# DRAWING OF THE WIRE OF LOW-CARBON STEEL: PLASTICITY RESOURCE, OPTIMAL REDUCTION, STRUCTURE, PROPERTIES.

A. Zavdoveev[1,2], Ya. Beygelzimer[1], E. Pashinskaya[1], V. Grishaev[1], A. Maksakova[1].

[1] Donetsk institute for physics and engineering NAS of Ukraine.
[2] Paton Electric Welding Institute NAS of Ukraine.

**Abstract**

The work considers the effect of deformation on the exhaustion of the plasticity resourceof steel 45 after the drawing deformation. The results of theoretical and experimental studies of damage accumulation are listed. A possibility of employment of a scanning electron microscope to observe submicropores is demonstrated.

**Introduction**

The problem on nucleation, growth and propagation of cracks seems to have no direct relation to the drawing of well-deformed materials, namely, steel and iron. The supposed probable formation of cracks in the course of processing seems to be rejected due to the fact that the processing enhances the properties of a steel wire (an increase in the number of inflections, kinkings, necking-down). But a number of the wire properties definitely indicate to the possible emergence of micro- or submicrocracks affecting the plastic properties and some operation characteristics.

Cold-worked state of the carbon steel and iron is associated with a peculiarity that is not usually considered in respect of the mechanical properties of steel. The point in question is a density decrease in the course of cold deformation of steel. It is known that dislocations and vacancies increase the volume of the cold-worked material [1, 2], but the simplest calculation indicates that even the ultimate density of dislocations and vacations cannot increase the volume by more than 0.1%, whereas the registered values are of 0.5-0.6 and even 1.0% of the material [3].

In the course of nucleation and evolution of micro-discontinuities under plastic deformation, simultaneous healing of the defects occurs with the intensity controlled by the hydrostatic pressure and the temperature increase. Energetically unstable micro-discontinuities generated under metal working can be healed by recovery annealing, contrary to the energetically stable ones. Proper planning of deformation provides annealing before the micro discontinuities become energetically stable. The theory of deformation of metals allows accounting for this fact.

Now the most wide-spread approach is based on the assumption about the existence of some state of the material that can be quantitative described by the damage. The fracture occurs when the damage reaches the maximum.

The present paper contains the calculation and the experiments for the evaluation of the damage accumulation in the course of cold drawing of steel 45.

# EXHAUSTION OF THE PLASTICITY RESOURCE OF METALS UNDER DRAWING

All kinds of the defects are generated and accumulated in the material in the course of plastic deformation. To describe this phenomenon, the postulate was suggested and developed in [11-23] that there exists a macro-object called damage that is the quantitative measure of the microfracture of the metal under deformation. The last phenomenon precedes the macrofracture immediately.

According to the principle of macroscopic definability by Ilyushin [24], the magnitude of the damage is supposed to be univocally defined by the loading process, being presented in the form of a functional of the process.

The criterion of the macroscopic fracture is written as a condition when a certain degree of the damage achieves the critical value.

The works reporting this approach to the modeling of the fracture differ in the choice of the object treated as the "damage" (a scalar as in [11-15, 19-23], or a tensor as in [20-22]), in the form of the functional expression and the measure of the damage entering the criterion of the fracture.

Now the metal working researchers mostly use the criteria based on the scalar measure of the damage [11-19, 20-24]. The point is that the use of tensor variables must be based on more perfect theories of plasticity that have not been developed for now (there are no sufficient theoretical and experimental reasons for this kind of theories). The main idea of the approach realized in the listed works is as follows.

The measure of the damage is a scalar variable $\varepsilon$ called cleavage in the basic work by V.L. Kholmogorov [11] and softening in further publications on the topic. It is commonly supposed that an increase in the softening is proportional to the increment of the shear strain $d\Lambda$:

$$d\varepsilon = \alpha * d\Lambda , \qquad (1)$$

where $\alpha$ is the factor determining the intensity of accumulation and evolution of microcracks.

It is assumed that the formation of a microscopic crack occurs at the moment when the softening reaches the critical value $\varepsilon_{\kappa p}$. The degree of the shear strain applied to the representative volume of the metal before the critical softening was achieved is called plasticity and labeled by $\Lambda_p$. Supposing $\alpha = $ const when the metal is deformed, it follows from (1) that $\varepsilon_{\kappa p} = \alpha \Lambda_p$. Dividing (1) by the last relation yields

$$d\psi = \frac{d\Lambda}{\Lambda_p} , \qquad (2)$$

where $d\psi = \frac{d\varepsilon}{\varepsilon_{kp}}$, and $\psi$ is called the degree of exhaustion of the plasticity resourse.

It results from (2) that

$$y = \int_0^t \frac{H dt}{L_p}, \qquad (3)$$

where $H dt = dL$, $H$ is the intensity of the velocity of the shear strain.
The condition of the deformation without fracture is

$$y = \int_0^t \frac{H dt}{L_p} < 1, \qquad (4)$$

and the fracture condition is

$$\psi = \int_0^t \frac{H dt}{\Lambda_p} = 1. \qquad (5)$$

The results of the plasticity tests under the proportional load with sufficient accuracy can be presented by parametric dependences describing the relation of the limit shear strain $\Lambda_p$ to the parameters of the stressed state $\frac{\sigma}{T}$ ($\sigma$ is the hydrostatic stress, T is the intensity of the tangent stresses); $\mu_\sigma$ is the Nadai-Lode parameter; as well as the intensity of the velocity of the shear strain $H$ and the temperature $\theta$:

$$\Lambda_p = \Lambda_p\left(\frac{\sigma}{T}; \mu_\sigma; H; \theta\right) \qquad (6)$$

In [11-24], the dependences (6) are listed for a number of metals and alloys.

Relation (3) allows evaluation of the degree of the exhaustion of the plasticity resourcewhen the temperature, the stressed state and the deformed one are known (in other words, $\theta$, $\frac{\sigma}{T}$, $\mu_\sigma$ and $H$ along the trajectory of the particle motion in the deformed metal are in disposal), in the case of the known parameter dependence of the metal plasticity (6). The analysis of the results of calculations permits making a conclusion about possible fracture of the metal during the process and fixation of the areas with the highest fracture probability in the billet.

We think that the most important application of the theory of deformability in metal working is not only the forecast of the fracture moment according to criterion (5), but also the estimation of the damage of the metal by the evaluation of $\psi$. This approach allows predicting of the parameters of the quality of the billets and articles related to the damage and proper selection of the modes of recovery anneals. For this purpose, the theory of deformability develops conceptions of the anneal effect on the degree of the exhaustion of the plasticity resource[13,20-24].

It was mentioned above that the microdiscontinuities can be stable or unstable from the energetic viewpoint. The first-type discontinuities can be healed in the course of the recovery anneals, contrary to the second-type ones. This fact means that there is a critical parameter $\psi$ that controls the efficiency of anneals. At $\psi < \psi$, the microdiscontinuities can be totally healed and the undamaged metal structure can be recovered by the anneal; at $\psi > \psi$, only partial elimination of the damage is possible. The magnitude of $\psi$ depends on the deformed metal, the concrete data are reported in [20-24]. The characteristic range is $\psi = 0.2 \div 0.4$.

According to [20-24], there is the second critical value $\psi$ associated with sharp decrease of the recovery of the plasticity resourse. When the unhealed micropores appear above $\psi$, unhealed microcracks arise above $\psi$. The characteristic range is $\psi = 0.5 \div 0.7$.

Criterion (4) agrees to the experiment under a simple load only (or a load close to the simple one). The processes of a complex load require more perfect criteria because the axis of the principal strain velocities are rotated through substantial angles with respect to the metal particles (see [16-19,20,23]). They account for the experimental fact of enhanced metal plasticity after sign-alternating deformation.

Kinetic equation (1) does not consider the processes of microdiscontinuity healing. The only indirect reference is expression of the ultimate plasticity (6). This demerit results in restriction of the area of adequacy of the fracture model and a number of necessary calibration experiments. This fact was several times pointed out by V.L.Kholmogorov. In a fundamental work [11], he made the first steps to explicit account of the healing of microdiscontinuities, but the work had not been finished.

Recently the Ural school of metal working proceeded with this research area. In [23], a generalization of the kinetic equations of the phenomenological theory of fracture has been made in the case when the plastic deformation is combined with thermal treatment of a metal. In [23], an attempt was made to formulate the problem of deformation an the fracture of the metal in the course of metal working with account of the effect of fracture on the stress-strain state of the metal. In this work, the approach to the study of the fracture under metal working described above is aligned with another approach developed recently. The necessity of the search for new approaches is determined by two reasons, at least. The first reason is that the magnitude of the damage can not be measured directly. It is evaluated only in the case of the macroscopic fracture of the material. The second reason is that the criteria [15-23] do not described the fracture associated with the strain localization.

The solution of the mentioned problems is possible within the frameworks of the fracture modeling reported in [4, 12, 24] only. The point is that the magnitude of the relative porosity is assumed to be a measure of microfracture, and the related kinetic equation can be written on the basis of one or another consideration. As shown above, this approach is well substantiated because the fracture is associated with the formation of microdiscontinuities determining the value of the porosity [24].

Contrary to the damage, the porosity can be directly measured, and the first abovementioned problem can be solved.

The works performed within the frameworks of this approach differ in the methods of derivation of the kinetic equations for the porosity.

We believe that the studies where the kinetic equations are derived from the basic relations for a deformed material [12, 20-24], are of the highest interest. The fact is that the localization of the plastic strain and the related fracture can be studied within the frames of the research. So, the second problem is solved.

The described approach is very promising, being sufficiently agreed with the physical conceptions of the fracture under metal working.

We shall apply the simplest variant of the theory of deformability to calculation of the plasticity resourceexhaustion when a wire of steel 45 is drawn.

In Fig.1 the plasticity diagram of the steel [25] is presented.

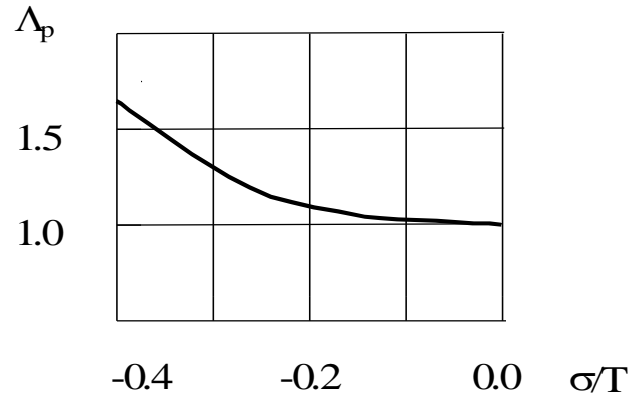

Fig.1 – Plasticity diagram of steel 45 (see [25])

In the same paper, a dependence approximating the experimental curve was presented:

$$\Lambda_p = \exp(-1.11 \frac{\sigma}{T}) \qquad (7)$$

In the course of drawing, the exponent $\sigma T$ is measured along the deformation zone. According to [38], in the first approximation one can accept that

$$\sigma T = 1.73(1 + m \cdot ctg\alpha t \cdot \ln \frac{f_0}{f} - 1.15 \qquad (8)$$

where m is the friction coefficient, $\alpha$ is the half-angle of the die cone, $f_0$ is the sectional area of the wire before drawing, $f$ is the current sectional area of the deformation zone.

Relation (8) shows that $\sigma T$ under drawing (without back-tension) ranges from $(\sigma T)_0 = -1.15$ at the inlet of the deformation zone ($f=f_0$) to $(\sigma T)_1 = 1.73(1 + m \cdot ctg\alpha t \cdot \ln \frac{f_0}{f_1} - 1.15$ at the outlet ($f=f_1$, where $f_1$ is the sectional area of the wire after drawing).

In the first approximation [38], the increment of shear deformation after drawing can be determined by

$$d\Lambda = -\frac{df}{f} \qquad (9)$$

If we substitute (7)-(9) in (3), we get the following expression of the exhaustion of the plasticity resourceafter a pass through the die:

$$\psi = 0.252 \frac{\lambda^{1.92(1+m \cdot ctg\alpha)} - 1}{1 + m \cdot ctg\alpha}, \quad (10)$$

where $\lambda = \frac{f_0}{f_1}$ is the elongation ratio of the die.

When passing to single draftings $\varphi$ and accounting of $\lambda = \frac{1}{1-\varphi}$, we get a formula describing the exhaustion of the plasticity resourceafter the drawing of a steel wire

$$\psi = 0.252 \frac{\left(\frac{1}{1-\varphi}\right)^{1.92(1+m \cdot ctg\alpha t}-1}{1+m \cdot ctg\alpha} \quad (11)$$

where $\lambda = \frac{f_0}{f_1}$ is the elongation ratio of the die.

Formula (11) can be applied to the calculation of the exhaustion of the plasticity resourceafter the sequential drawing.

The employment of the formula allows estimation of the value of single draftings when the recovery of the damaged metal structure due to anneal is still possible. In other words, there appears a possibility of rational control of the anneal.

The simplest model we have developed allows testing of the effect of disintegration of deformation on the exhaustion of the plasticity resourse. Suppose it is necessary to obtain some total drafting $\varphi\Sigma$ per two passes. We study the relation of the total exhaustion of the plasticity resource $\psi\Sigma$ and the pass distribution of the drafting. The value of $\psi\Sigma$ is determined by the formula valid for the monotonic deformation:

$$\psi\Sigma = \psi 1 + \psi 2 \quad (12)$$

where $\psi 1$ and $\psi 2$ are the exhaustion of the plasticity resourcefor the first and the second passes, respectively.

It can be easily shown

$$\varphi 2 = \frac{\varphi\Sigma - \varphi 1}{1 - \varphi 1} \quad (13)$$

where $\varphi 1$ and $\varphi 2$ is the drafting for the first and the second passes, respectively.

In Fig.2, the relation of $\psi\Sigma$ and $\varphi 1$ is illustrated.

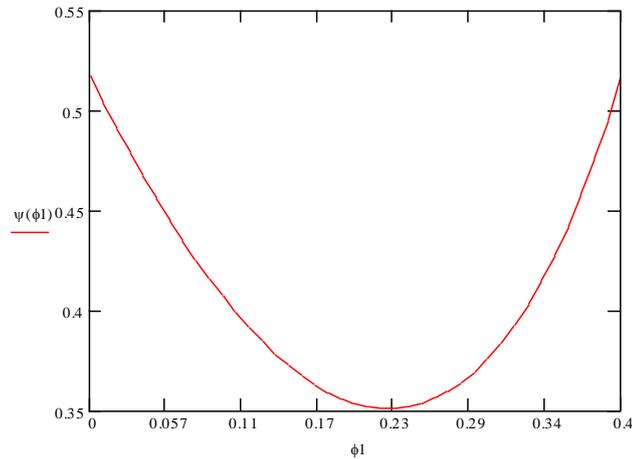

Fig.2 – Relation of the total exhaustion of the plasticity resource and the drafting during the first pass ($\varphi^\Sigma$ =0.4, α=8⁰, m=0.05), calculated by (11).

Fig.2 shows that redistribution of the strain over the passes can substantially reduce the exhaustion of the plasticity resource. At $\varphi^\Sigma$ =0.4, the most preferable variant of the drafting redistribution is $\varphi^1$ =0.23 and $\varphi^2$ =0.22. In this case, almost the whole damage accumulated by the metal can be healed by the recovery anneal ($\psi^{\Sigma}=0.35=\psi$), contrary to the single drafting of 40% ($\psi^{\Sigma}=0.52>\psi$ =0.35).

In Fig.3, single drafting dependence of the exhaustion of the plasticity resource of steel 45 is illustrated, being calculated by (11).

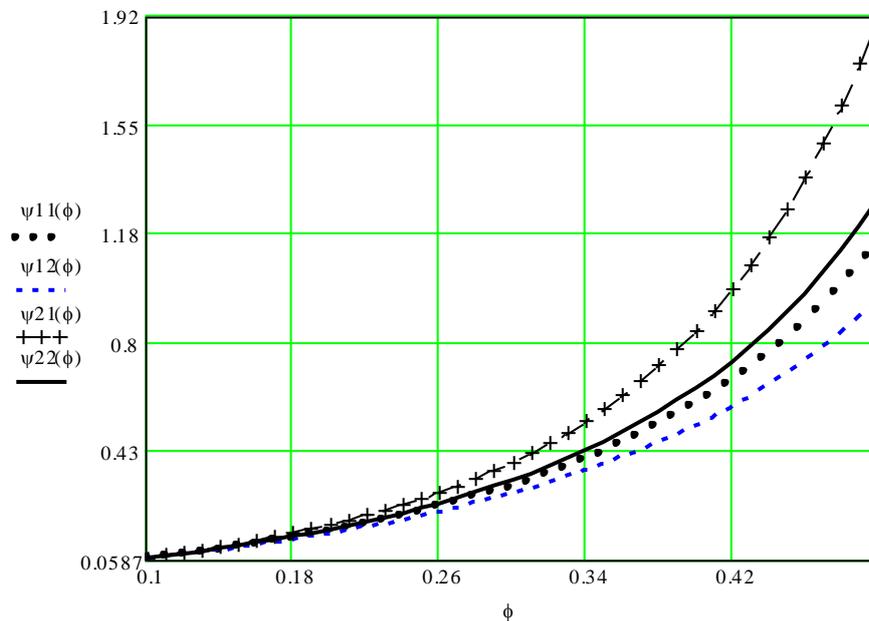

Fig.3. Single drafting dependence of the exhaustion of the plasticity resource of steel 45: α = 5⁰, m=0.05 (ψ11); α = 5⁰, m=0.10 (ψ21); α = 8⁰, m=0.05 (ψ12); α = 8⁰, m=0.10 (ψ22);

Fig.3 shows that variant 12 is the most preferable with respect to plasticity (α = 8⁰, m=0.05), and variant 21 is the least preferable (α = 5⁰, m=0.10). By using (11), we can evaluate single draftings when the recovery of the damaged structure by anneal stays possible. The critical value of steel 45 is $\psi$ =0.35. It follows from (11) that this value is associated with the maximum drafting 33% (in the case of the most preferable variant 12).

An enhancement of the metal plasticity due to disintegration of the metal is a known fact. The example mentioned above can not be considered to be a demonstration of the abilities of the theory of deformability. This is rather a short test illustrating validity of the theory conclusions in the case. It becomes really useful when dealing with multioperational processes of metal working including intermediate anneals (for instance, multi-pass drawing). A number of possible variants of the technology arise here and the most rational method should be selected with respect to a number of the parameters. In these cases, adequate rheological models of the deformed materials should be employed as well as powerful methods of calculation of the stress-strain state and modern variants of the theory of deformability. At the same time, the technique should be developed with the use of the effective approaches of the optimum design: dynamic programming, methods of multicriteria optimization, sensitivity analysis etc. (see [12, 38, 39]).

This method allows forecasting the exhaustion of the plasticity resourcein the course of drawing and putting the achievements of the theory of deformabilty into practice when designing the drawing routes that can provide high quality of the wire with respect to the plasticity parameters.

The experimental verification of the obtained formules have been finished now. In the course of the experiment, the value of the accumulated strain reached $\psi$ = 0,75. The formation of micropores and microcracks was registered in the metal structure during the drawing of the steel wire.

EXPERIMENTAL PROCEDURE

A number of works demonstrate the formation of micropores in the course of drawing [1,2]. The method of small-angle X-ray scattering allowed indirect registration of nanopores of 20..30 nm in size after ECAP processing [3], i.e. under uniform compression that seems supposed to be preventing the generation of microdiscontinuities. However the evolution of the metal structure in the course of SPD is that the strong reduction of the grains is accompanied by a substantial misorientatiion of the grain boundaries (over $15^0$), the boundaries are transformed into high-angle ones. As a result, the boundaries become amorphous, weakened and incoherent; their volume fraction increases and becomes comparable with the volume fracture of the very grains. This fact facilitates the rotation of the grains and modifies the mechanism of the plastic deformation, terminating the strengthening of the material. Probably, the incoherence of high-angle boundaries is provided by the accumulation of nanopores within them. Even the presence of microdiscontinuities of small size affects the mechanical properties of the materials. Thus, the direct observation of nanoporosity becomes very actual because of possible verification of the conceptions about the reconstruction of the metal structure under SPD.

Traditional testing of the structure by an electron microscsope implies processing of the surface by smoothing and polishing. As a result, the pores become invisible, being closed with the processing products. There is a known method of the study of the fracture surface formed by a brittle crack that is aimed at detection of the pores of micron size [4]. Before the fracture, the sample is cooled in liquid nitrogen up to the state of coldbrittleness. After that, a starting microcrack is induced and wedged up to the fracture .

We have suggested another simpler method of the sample fracture. A cooled sample is fractured by the three-point bend (Fig.4). A concentrator in the form of a sharp cut is preliminary applied to the surface. A turning workstation or a planning machine is used to obtain a sample of the round or rectangular section, respectively. The labels are made in order to place the sample on the end supports. The labels are located symmetrically about the concentrator and the distance to the concentrator is no more than 2,5 heights of the sample section. The load is applied to the centre in the plane of the concentrator. At these conditions, the crack propagates strictly in the plane perpendicular to the axis of the sample [5,6].

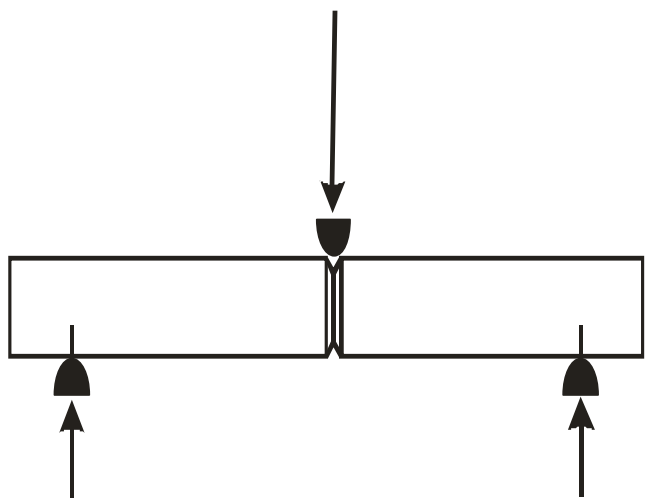

Fig. 4 — Three-point scheme of the fracture

Depending on the aim of the experiments, the samples were made of different materials of varied cross-section form. First of all, it was necessary to be sure that an electron microscope allows registration of pores of hundreds and tens of nanometers on the surface of the brittle fracture. For this purpose, construction steel 45 was used after thermal treatment up to the state of the lowest plasticity (quenching only). The sample (Fig. 5) was of 14 mm in diameter, with a ring concentrator of 2 mm in depth. The length of the load levers was 30 mm. The fracture was performed by an impact by the three-point scheme.

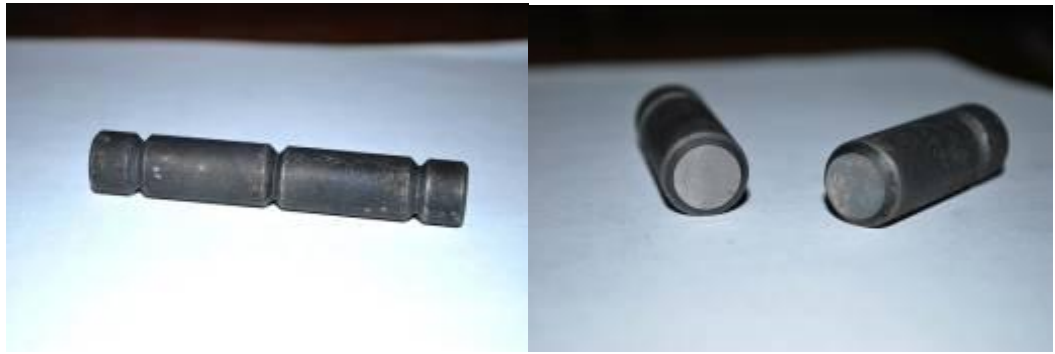

Fig. 5 – The sample before and after the fracture

Cold deformation of steel 45 was performed by multi-pass drawing in order to achieve the strain about unity. The drawing was made at a drawing bench along the route (8-7.6-6-5.6-5-4). The analysis of the hardness density anbd the structure was made with using the electronic scales Shimadzu 200, the durometer B5, the microscope Axiovert 40MAT. The fracture surface was analysed by the scanning electron microscope JSM-6490LV (JEOL, Japan) in the SEI mode.

RESULTS AND DISCUSSION.

The fracture surface was analysed by the scanning electron microscope JSM-6490LV (JEOL, Japan) in the SEI mode. The images of the surface structure of steel 45 at different magnifications (Fig.6).

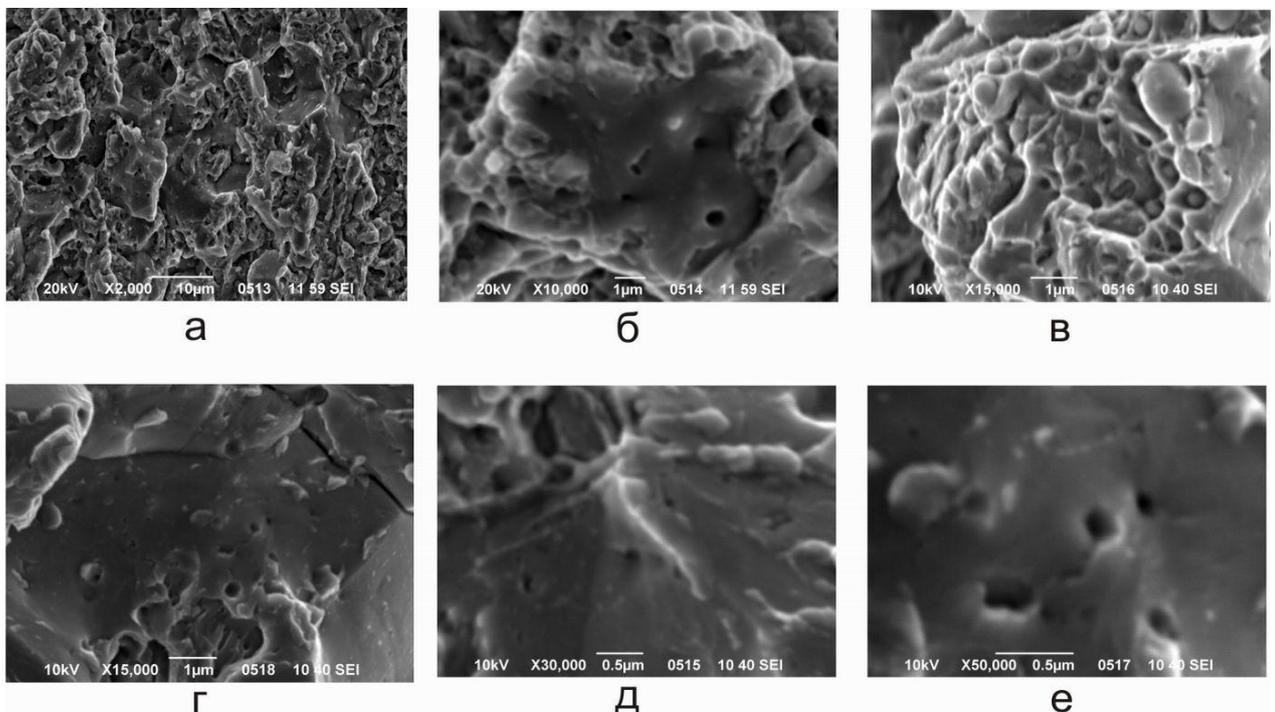

Fig. 6 – Microphotos of the fracture of steel 45

The mode of the microscope work allows to see the fracture surface due to the secondary electron emission of the surface details. The inclined convex and concave areas of the surface are registered as bright and dark areas, respectively. We see a number of small bright convexities with the corresponding concavities of the same size that can be treated as dimples. These objects are complementary, i.e. the convexities of the first fracture surface correspond to the concavities of the secodnd part of the fractured sample.

At the same time, dark spots of 0,1 μm and lower are found, that are looking like extended recesses, not dimples. We can not register the corresponding extended convexities because they would be unavoidably broken in the course of the fracture. Dark spots of this type are usually interpreted as pores, including not round ones [10].

Finally, small nanoobjects can appear to be some non-metallic inclusions. To be certain, it is sufficient to test the chemical composition of a spot. The data of the SEM analysis demonstrate that the basic element of the tested areas of the surface is iron (over 90%). This fact permits to interprete dark areas of hundreds and tens nanometers in size as pores.

Thus, the pores of submicro- and nanosize can be registered on the surface of a brittle fracture. Naturally, the porosity can be described as a surface porosity, i.e. the ratio of the dark spots treated as pores to the tested surface area. Extrapolation to the volume porosity provides no additional information.

The analysis of microstructures was performed with using the metallographic microscope Axiovert 40 MAT. The microstructure of the sample before drawing is ferrite-perlite structure with the perlite content about 50 %. In the initial sample, the ferrire grain size was about 20 μm. After the deformation, the redistribution of the perlite component was registered as well as the emergence of granular perlite. The perlite component has also been redistributed and reduced. Besides, the transient areas between the lamellar and granular perlite have appeared.

The analysis of the microstructure tests has shown that the drawing results in work hardening, i.e the reduction of the structure components. In particular, the initial average ferrite grain size was about 20 μm, contrary to 10 μm after the deformation (Fig. 7).

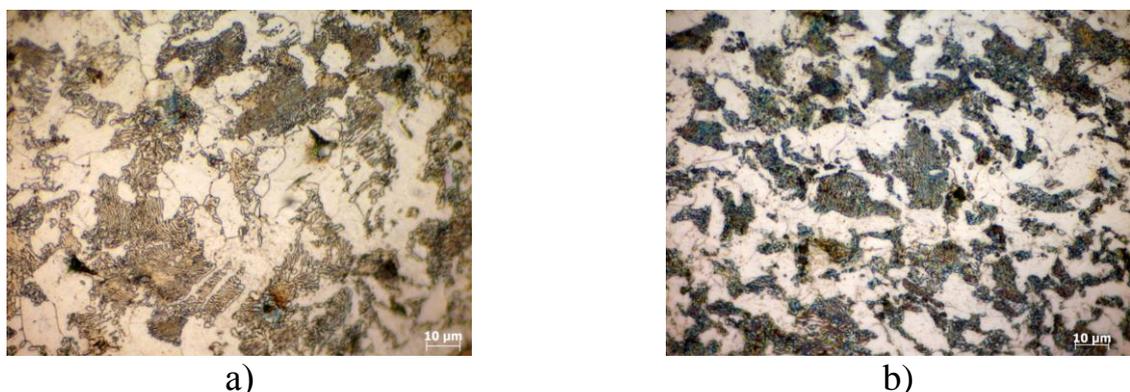

a)        b)

Fig. 7 — Microstructure of steel 45: a) the initial annealed steel, $800^0$C 1 h., b) after the drawing, $e$=0.98.

Besides, an increase in strain results in increase in the strength of the samples due to work hardening.

The dependences of these parameters ane presented in Figs. 8 and 9.

The figures illustrate strong correlation between the density and the hardness of the processed material that is strain-dependent. In particular, an increase in strain is accompanied by a hardness increase and a density reduction. An analogous behavior is observed in the relation of the hardness, the density and the exhaustion of the plasticity resourse.

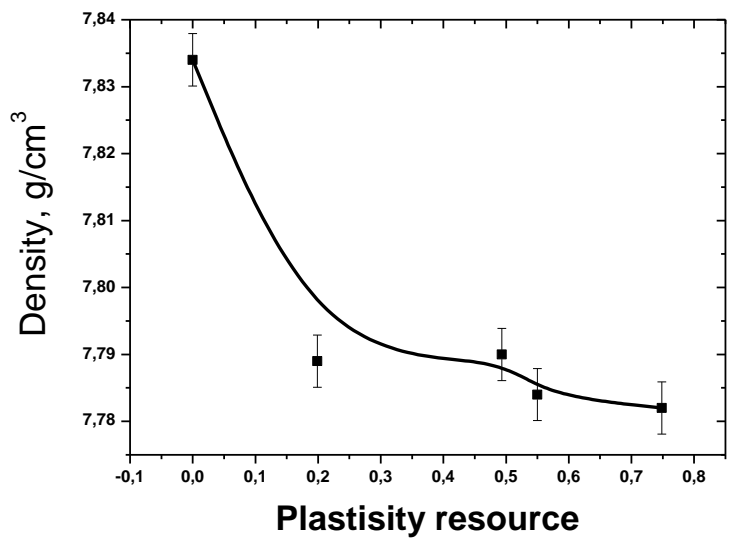

Fig. 8 – Dependence of density vs. plasticity resource.

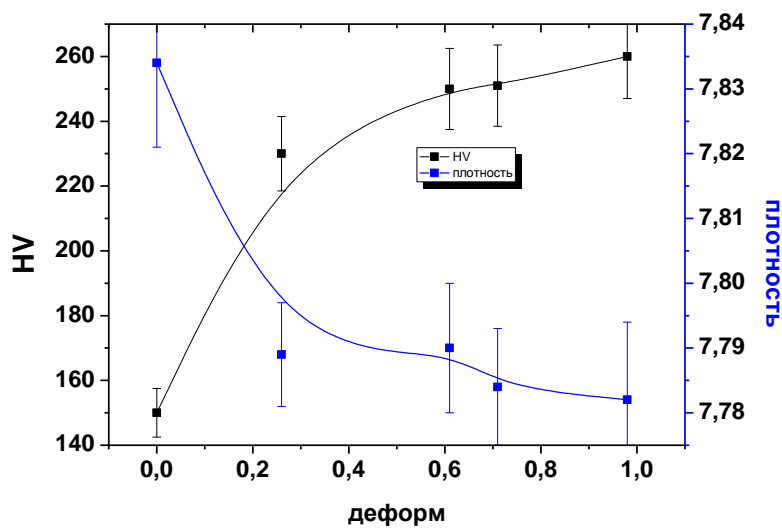

Fig. 9 – Dependence of hardness and density vs. deformation.

The density decrement was about 1% during the whole deformation course. The decompaction of the material can be explained by the formation of pores and microcracks. In this context, to confirm the existence of micropores and cracks, brittle fracture of steel was performed before and after the deformation. It is explicitly seen in Figs. 10 and 11, that the fracture is brittle. The original sample has been broken by the mechanism of brittle fracture, so the micropores and the cracks are almost absent. The deformed sample demonstrates multiple formation of pores and cracks.

Thus, it has been demonstrated in the work that microdefects are generated within the metal under drawing. The results are the density reduction and the enhancement of the plastitity resourse.

It has been shown theoretically and experimentally that the exhaustion of the plasticity resourcein the course of drawing is related to the accumulation of pores. The coalescence of the pores results in formation of the cracks and fracture of the material.

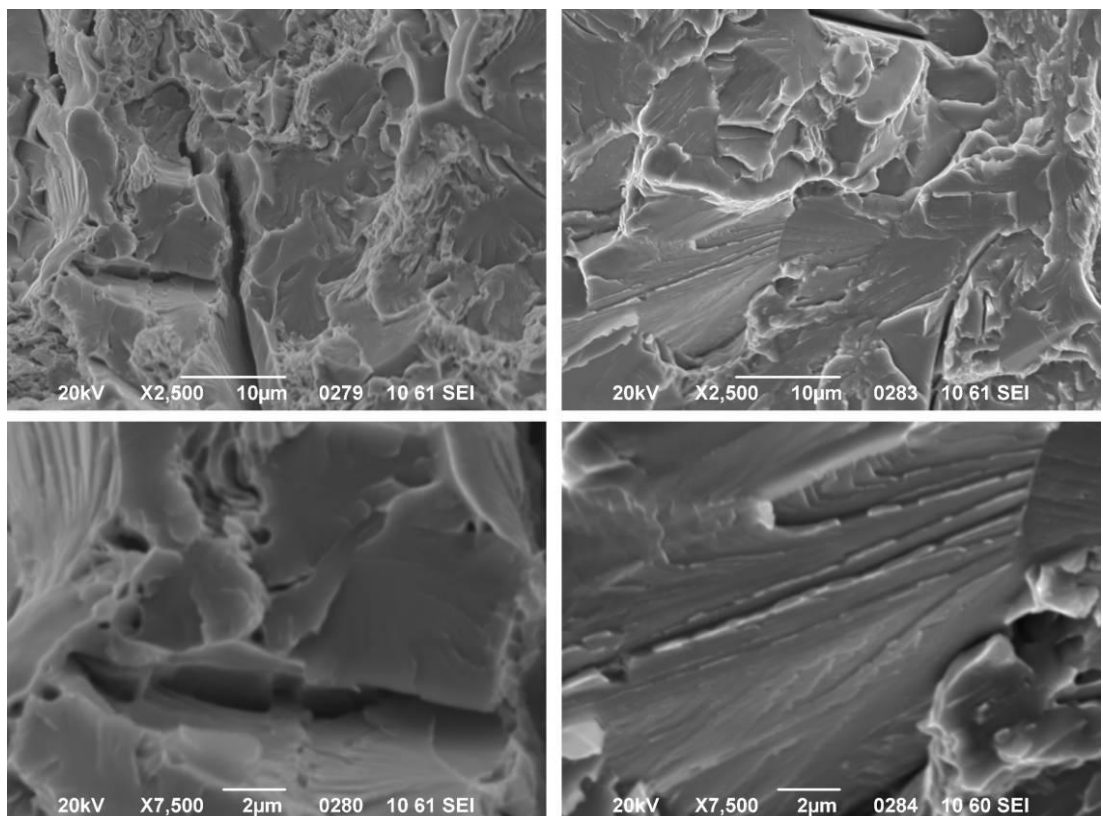

Fig. 10 – Fractograms of fracture surface: Initial – right, cold drawing $e$=0.98 – left.

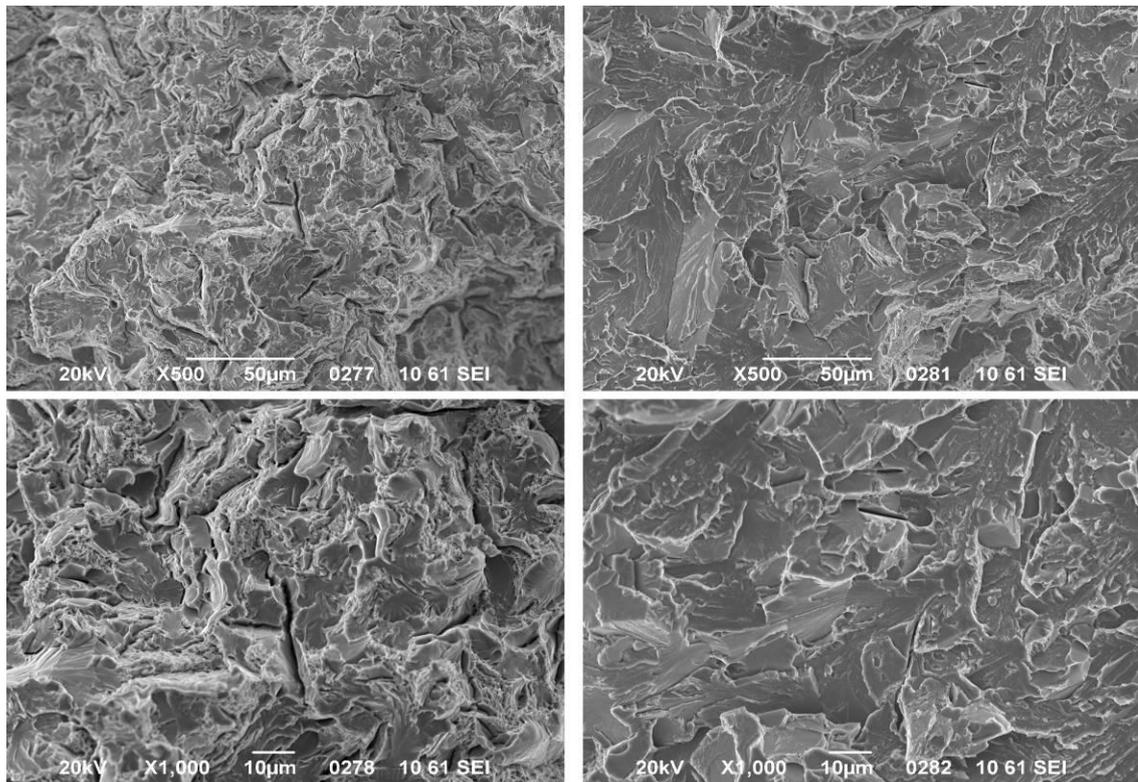
Fig. 11 - Fractograms of fracture surface: Initial – right, cold drawing *e*=0.98 – left.

CONCLUSIONS

In the present work, we have made an attempt to attract the attention of the researchers to the modern abilities of the metal working theory with respect to the forecast of the plastic characteristics of the metal. As an illustration, the simplest variant of the theory of deformability was considered. When analyzing the process of drawing theoretically and experimentally, the magnitudes of the ultimate single draftings and the exhaustion of the plasticity resourceafter the single and multiple drawing were evaluated. It has been shown that first the processes of the damage healing prevail after the single drafting, i.e. they are energetically unstable. When the strain becomes high, accumulation of the damage starts. The existing armoury allows employment of the theory of deformability to design the drawing routes that can provide high quality of the wire with respect to plasticity.

The theoretical analysis of structure imperfection calculated by the method of estimation of the exhaustion of the plasticity resourceshows that the increase in strain (i.e. the reduction of the diameter of the original billet) is associated with an accumulation of the structure imperfection that is expressed as an increase in the plasticity resourse. Additional studies performed by hydrosatic weighing have also shown that the structure imperfection is accumulated as the strain increases.

The pores of submicro- and nanosize can be observed on the brittle fracture surface by a scanning electron microscope.